\documentclass[aps,prl,twocolumn,groupedaddress,amsmath,amssymb]{revtex4}

\usepackage{graphicx}
\usepackage{float}
\usepackage{subfigure}

\begin{document}


\title{Indium-Gallium Segregation in CuIn$_{x}$Ga$_{1-x}$Se$_2$: An ab initio based Monte Carlo Study}

\author{Christian D. R. Ludwig, Thomas Gruhn, Claudia Felser}

\affiliation{Institute of Inorganic and Analytical Chemistry, Johannes Gutenberg-Universit\"at Mainz, Germany}

\author{Tanja Schilling}

\affiliation{Institute of Physics, Johannes Gutenberg-Universit\"at Mainz, Germany \\
		Theory of Soft Condensed Matter, Universit\'e du Luxembourg, Luxembourg}

\author{Johannes Windeln}

\affiliation{IBM Mainz, Germany}

\author{Peter Kratzer}

\affiliation{Faculty of Physics, Universit\"at Duisburg-Essen, Germany}


\begin{abstract}

Thin-film solar cells with CuIn$_x$Ga$_{1-x}$Se$_2$ (CIGS) absorber are still far below their efficiency limit, although lab cells reach already 19.9\%. One important aspect is the homogeneity of the alloy. Large-scale simulations combining Monte Carlo and density functional calculations show that two phases coexist in thermal equilibrium below room temperature. Only at higher temperatures, CIGS becomes more and more a homogeneous alloy. A larger degree of inhomogeneity for Ga-rich CIGS persists over a wide temperature range, which may contribute to the low observed efficiency of Ga-rich CIGS solar cells.
\end{abstract}


\maketitle





As resources of fossil fuels are dwindling alternative sources of energy
gain increasing importance. The contribution of solar cells is continuously growing and a lot of effort has been made to improve the efficiency. During the past years chalcopyrites like 
CuIn$_{x}$Ga$_{1-x}$Se$_2$ (CIGS) have been shown to be promising absorber 
materials for thin-film solar cells with high efficiency and 
low production cost. Several crucial aspects of their operation have, 
however, not been completely understood yet. One question that is still unanswered is the influence of the In-Ga ratio on the cell efficiency. 
Pure CuInSe$_2$ has a band gap of 1.0~eV and CuGaSe$_2$ of 1.7~eV.
Aiming at an optimal band gap for the absorption of the solar spectrum, an alloy with about 70\% Ga should yield the highest efficiency. \cite{huang_effects_of_Ga}\cite{zunger_effects_of_Ga}. 
Experimentally, however, the best efficiencies have been reached with a much 
lower Ga content of only 30\% \cite{record_eff}. 
An explanation of this effect might lie in the inhomogeneity of the 
experimental samples. 
Knowledge on the impact of granularity on solar cell performance is still 
fragmentary, but inhomogeneities lead to band gap fluctuations, which most certainly have a detrimental 
effect on the cell efficiency \cite{eff_limit}\cite{bauer_pl1}. 
Several groups investigate experimentally the 
inhomogeneities in the In-Ga distribution for a fixed In-Ga ratio \cite{bauer_pl2}\cite{bauer_pl3}\cite{inhom_nano} or for varying In-Ga ratios \cite{bauer_pl1}\cite{pl_vary_Ga}. Photoluminescence measurements by G\"utay and Bauer indicate that a higher Ga content leads to larger inhomogeneities \cite{bauer_pl1}.
Even small fluctuations in the composition may unfavorably affect the electronic and optical properties \cite{eff_limit}. This might be an important factor that diminishes the efficiency of solar cells with high Ga content. 


In this letter we present a computer simulation 
study of the spatial distribution of In and Ga in CIGS. To make the calculations feasible we keep Cu and Se fixed at their respective lattice sites and neglect other defects.
It is common to tackle properties of semiconductors computationally by using 
density functional theory (DFT) for calculating their electronic structure. Up to now DFT-based calculations of CIGS compounds have only been carried out for small numbers of atoms and for the pure compounds CuInSe$_2$ and CuGaSe$_2$ \cite{zunger_electronic_structure_of}.
In this letter we discuss pattern formation on larger length scales at various temperatures. Studying large spatial inhomogeneities by ab initio methods in thermal equilibrium, however, requires a forbidding amount of CPU time, because one needs to sample hundreds of thousands of
configurations at large length scales.
We therefore used a hybrid method: using a cluster expansion (CE)
method \cite{ce_basics} we extracted interaction energies from electronic structure 
calculations and used these energies as input for Monte Carlo (MC)
simulations of large configurations of atoms. The calculational details are explained in the following.

The basic idea of the CE is to expand the formation energy $\Delta E_f$ of a configuration into energy contributions of ``cluster figures'' (single atoms, pairs, triples,...).

\begin{eqnarray}
  \Delta E_f & = & J_0 + \sum_{i} J_i s_{i} + \sum_{i<j} J_{ij} s_{i} s_{j} + ...
\end{eqnarray}

The indices $i$ and $j$ run over all lattice sites and $s_{m}$ is $-1$ for In and $+1$ for Ga on lattice site $m$.
Every figure is associated with a coefficient $J$ that gives the energy contribution of the specific figure. Detailed descriptions of the CE method can be found in \cite{ce_basics2} and \cite{ce_basics}. The coefficients of the expansion are fitted to ab-initio energies from electronic structure calculations.
Figures with a low value of $J$ are neglected to simplify the expression.

\begin{figure}
\begin{tabular}{ll}
a) &  b) \\

  \begin{minipage}{0.49\columnwidth}
    \includegraphics[width=\columnwidth,bb=0 0 638 650]{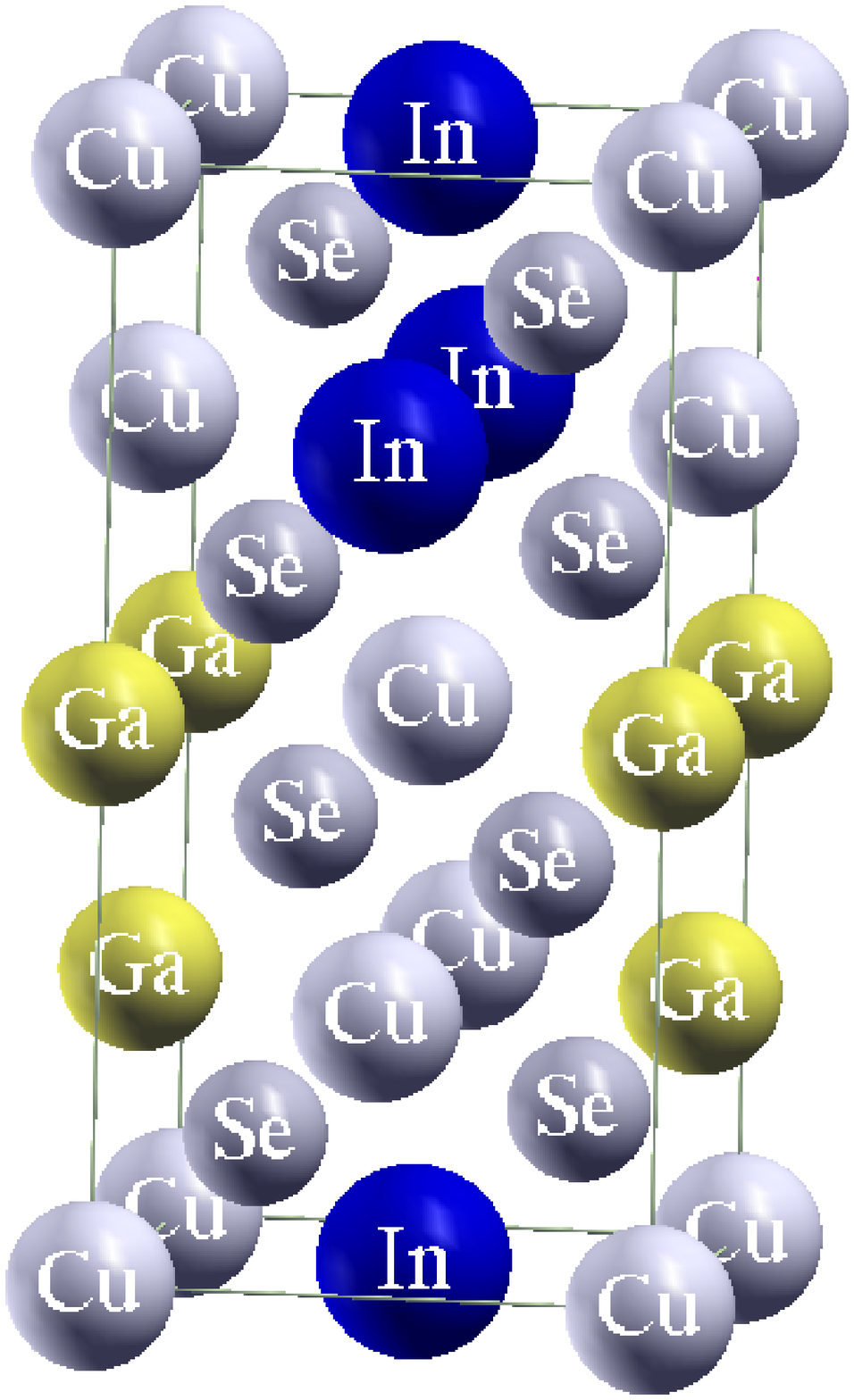}
  \end{minipage}
&
  \begin{minipage}{0.49\columnwidth}
    \includegraphics[width=\columnwidth,bb=0 0 638 650]{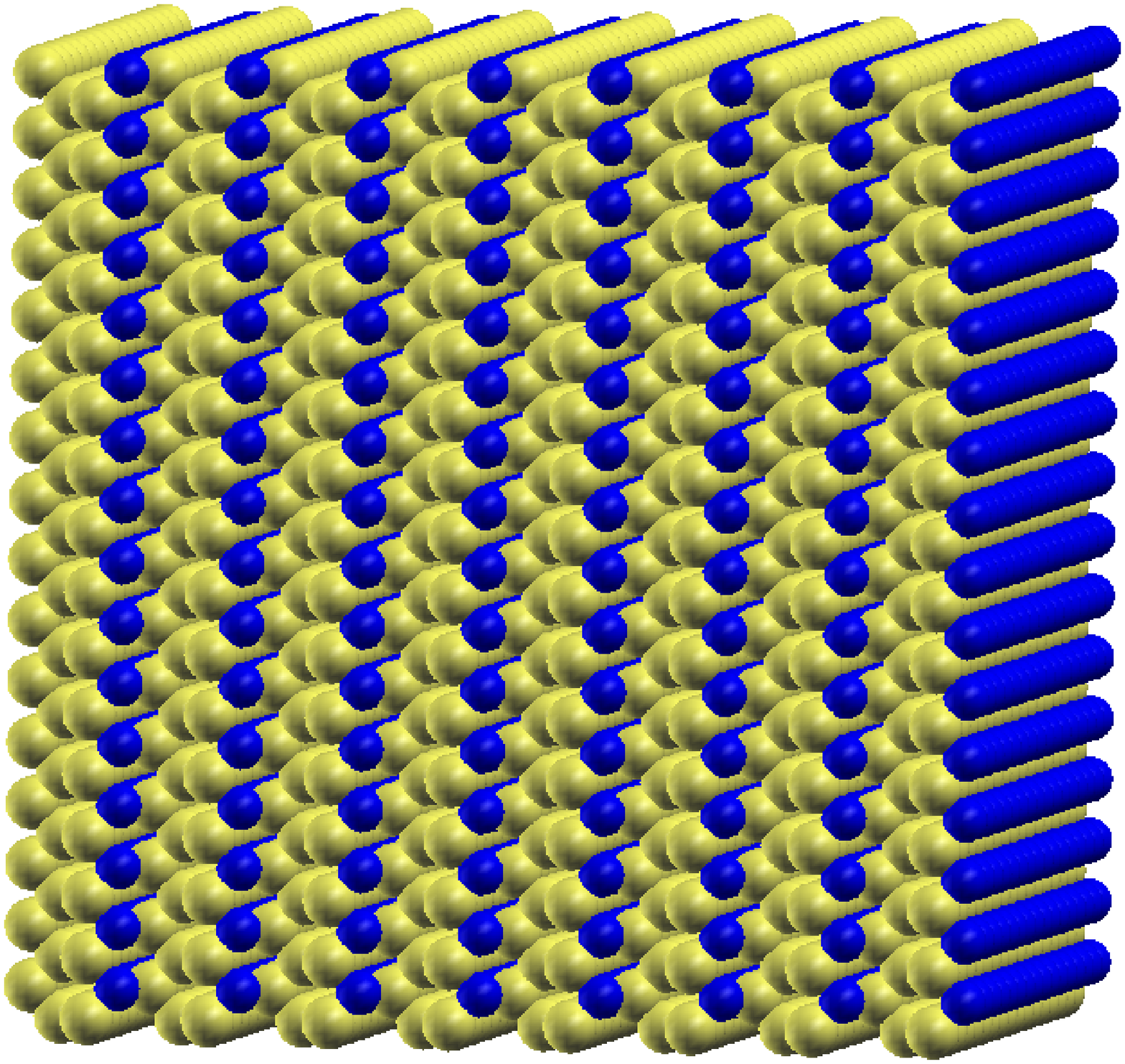}
  \end{minipage}

\smallskip \\

c) & d)\\

  \begin{minipage}{0.49\columnwidth}
    \includegraphics[width=\columnwidth,bb=0 0 638 650]{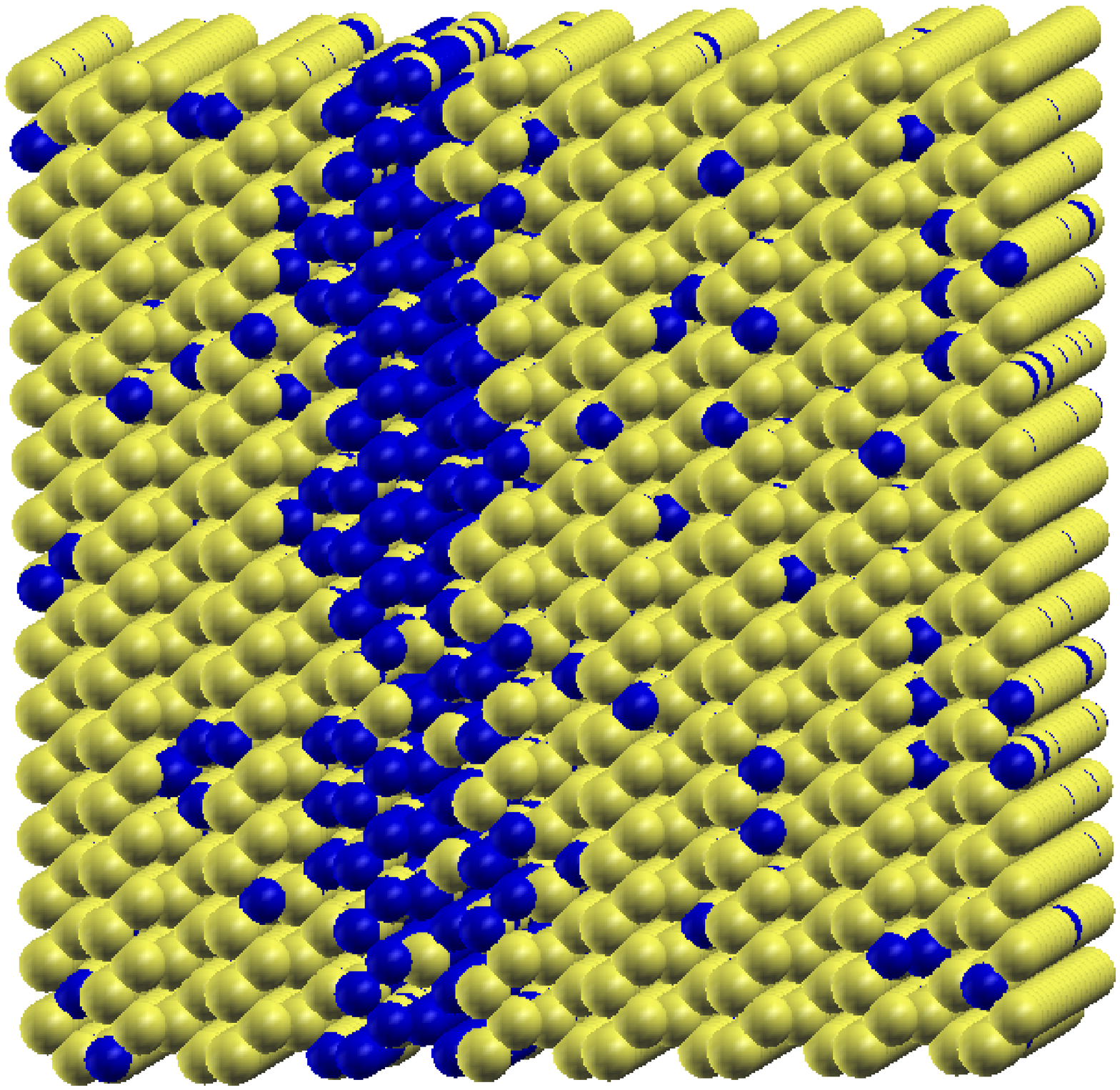}
  \end{minipage}
&
  \begin{minipage}{0.49\columnwidth}
    \includegraphics[width=\columnwidth,bb=0 0 638 650]{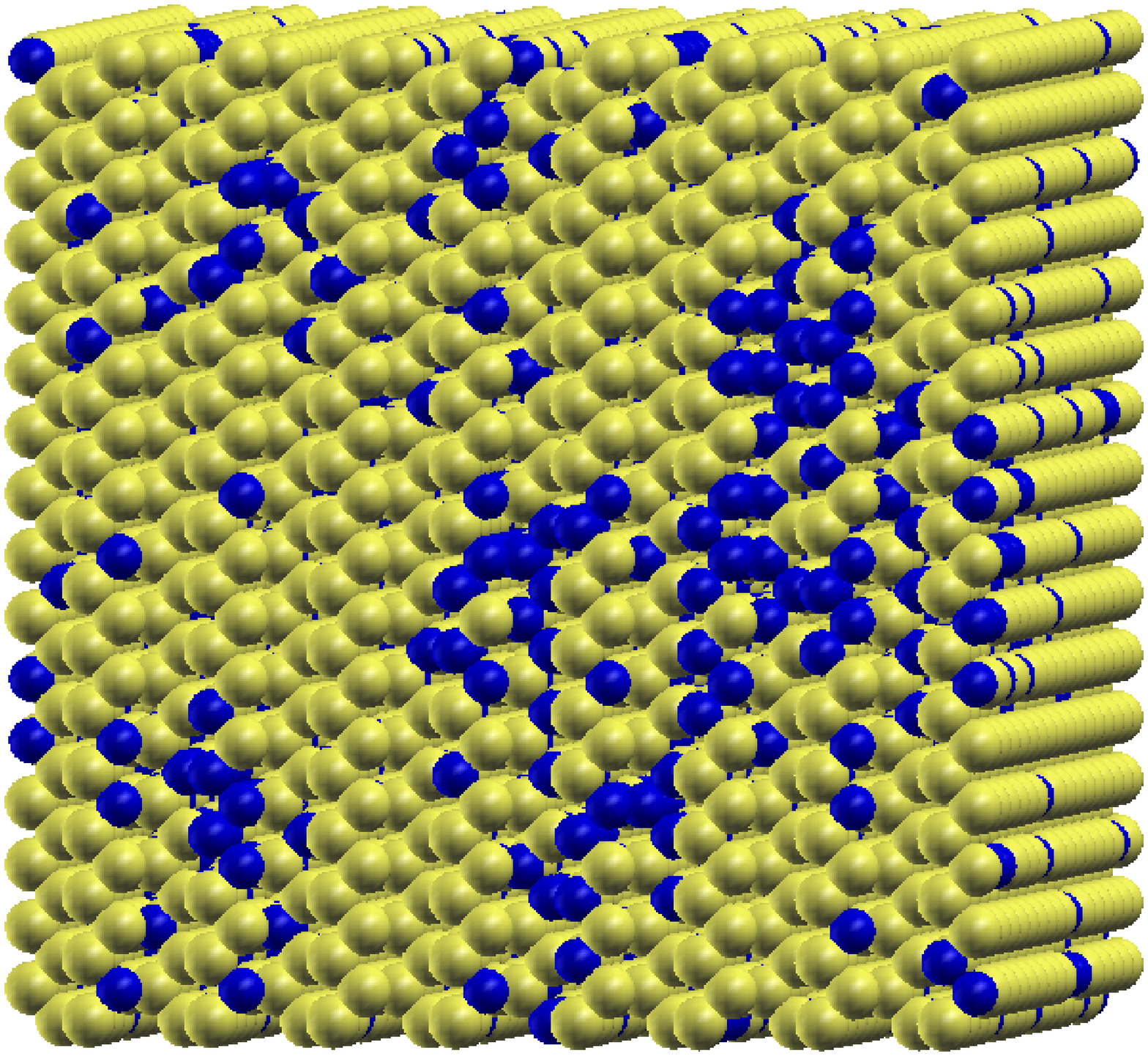}
  \end{minipage}
\end{tabular}
  \caption{a) CIGS unit cells with In and Ga atoms on Wyckoff position 4b. b) Snapshot of a system of periodic unit cells. c) Snapshot of the Ga-rich system at 30~meV / 348~K. d) Snapshot of the Ga-rich system at 35~meV / 406~K. Ga atoms are yellow, In atoms are blue. Cu and Se are not displayed in the snapshots and the size of the spheres is arbitrary.}
  \label{fig:snapshots}
\end{figure}

We calculated formation energies of 32 CuIn$_{x}$Ga$_{1-x}$Se$_2$ structures (space group I$\bar 4$2d, cf.~\cite{zunger_electronic_structure_of}) using the ab-initio electronic structure program \footnotesize ABINIT\normalsize \cite{abinit_note}\cite{abinit}. 
For the MC simulations it is necessary to vary the distribution of In and Ga (active atoms) on Wyckoff position 4b. Cu and Se atoms do not partake in the cluster expansion and are called ``spectator atoms''. The CIGS unit cell is shown in Fig. \ref{fig:snapshots}a). The generation of structures was automated using the Alloy Theoretic Automated Toolkit \footnotesize ATAT\normalsize \cite{atat}\cite{atat_userguide}. The structures include the CuInSe$_2$ and CuGaSe$_2$ unit cells and super-cells containing up to 32 atoms in total with In and Ga atoms distributed on Wyckoff position 4b. Trouiller-Martins-type pseudo-potentials were used with the generalized gradient approximation of Perdew, Burke and Ernzerhof \cite{pbe-gga} for exchange and correlation energies. The cut-off energy for the plane-waves was set to 70~hartree and a k-point grid of $3\times3\times3$ or bigger was used. The positions of all atoms were relaxed until the maximum force on atoms was less than $10^{-3}$~hartree/bohr and all three lattice parameters were relaxed until the stress was less than $10^{-5}$~hartree/{bohr$^3$}

\begin{figure}
\begin{tabular}{l}
a)\\
  \subfigure{
    \includegraphics[width=0.99\columnwidth,bb=50 50 410 302]{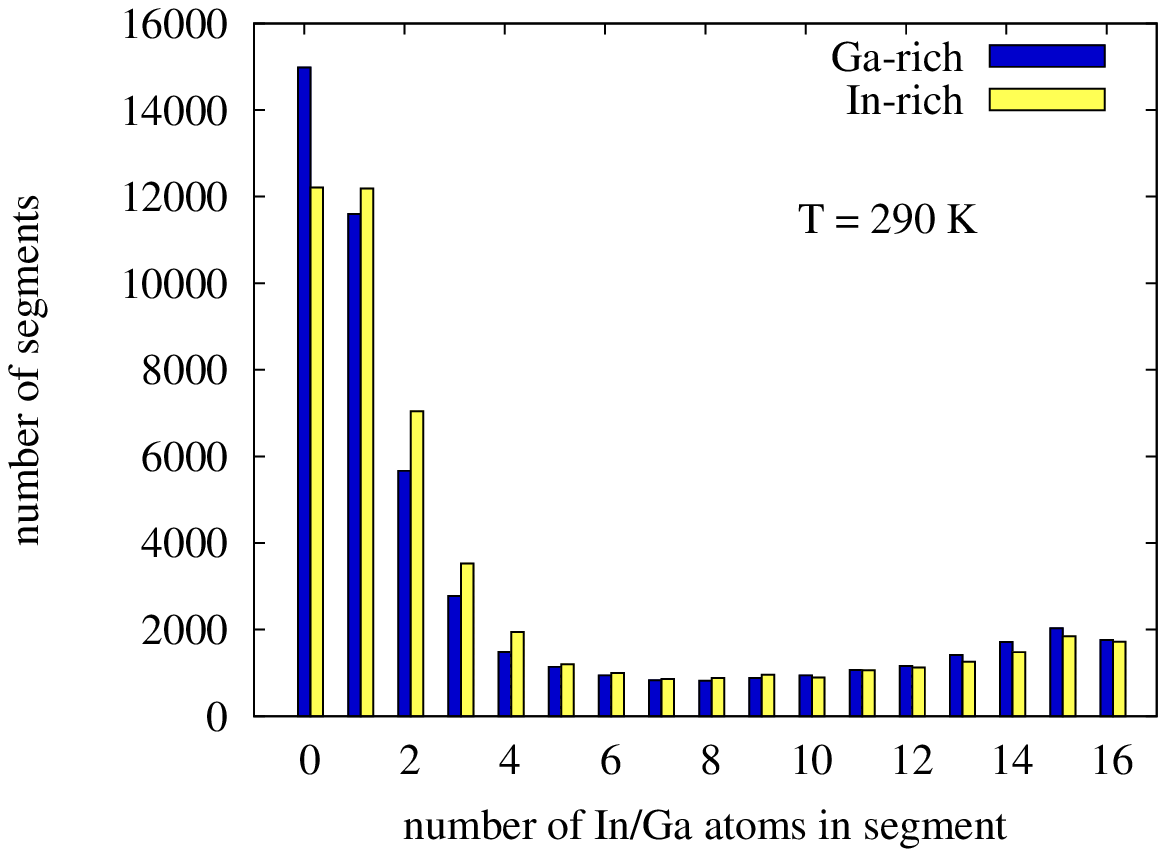}
  }\\
b)\\
  \subfigure{
    \includegraphics[width=0.99\columnwidth,bb=50 50 410 302]{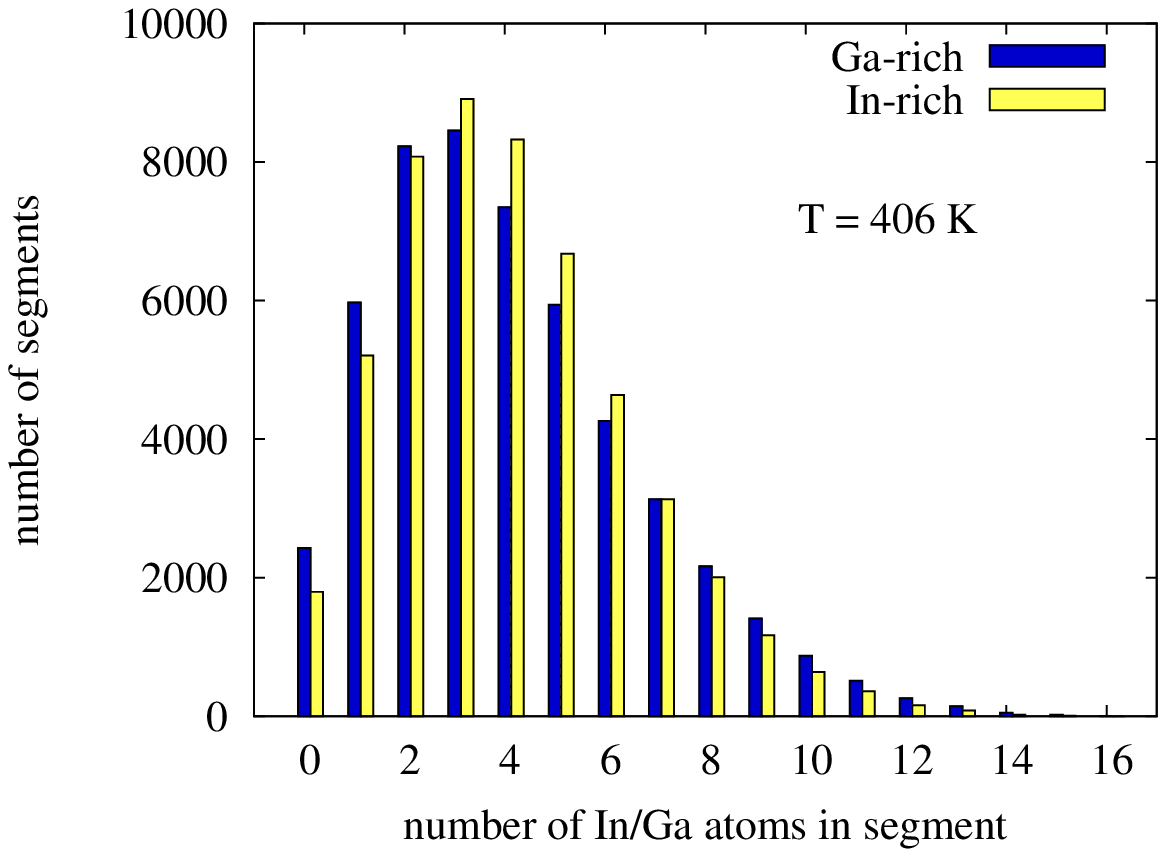}
  }
\end{tabular}
  \caption{Histograms showing the number of cubic segments in the simulation box that contain 1..16 atoms of a certain type. Blue is the distribution of In atoms in Ga-rich CIGS and yellow is the distribution of Ga atoms in In-rich CIGS at temperatures of a) 25~meV (290~K) and b) 35~meV (406~K). A perfectly ordered In-rich or Ga-rich system would have 4 Ga/In atoms in every segment. The histogram would have all entries in bin 4.}
  \label{fig:hist}
\end{figure}

\footnotesize ATAT \normalsize was used to construct a number of CEs. The effective cluster interactions (ECIs) which define the CEs were obtained by a least-squares fit to \footnotesize ABINIT \normalsize formation energies. From this number of CEs the optimal cluster figure set (with the lowest cross-validation score (CVS) \cite{kratzer_ce}) was chosen.
The CE with the lowest CVS of 1.3~meV contains one point figure, eight pair figures, two triple figures and two quadruple figures. The effects of constituent strain and volume deformation \cite{constituent_strain} were not taken into account, since we expect only a minor influence on the mixed state with fixed In:Ga ratio.

Canonical MC simulations were performed using the CE to calculate configurational energies. One MC move consists of exchanging the position of two active atoms (In/Ga). The simulation box contained 16$\times$16$\times$8 tetragonal CIGS unit cells. This translates into a cubic simulation box with 8192 active atoms. Simulations were run at several temperatures between 25~meV (290~K, approximately room temperature) and 75~meV (870~K, approximately the production temperature of CIGS thin-film solar cells) for $10^6$ MC sweeps. Typically, relaxation to the equilibrium state took less than $10^5$ MC sweeps.


For data analysis the simulation box was divided into cubic segments of 16 lattice sites. The number of In (Ga) atoms $b$ in each segment was counted and histograms were plotted. To have a measure for the inhomogeneity we computed the standard deviation $\sigma$ of these distributions. ($\sigma$ increases with increasing 
inhomogeneity.)

\begin{table}[h]
  \centering
  \caption{Standard deviation $\sigma$ for In-rich and Ga-rich CIGS and relative difference.}
  \begin{tabular}{c c c c}
    \hline
    \hline
    Temperature & $\sigma$ (In-rich) & $\sigma$ (Ga-rich) & $\frac{\sigma (Ga-rich)}{\sigma (In-rich)}-1$\\
    \hline
    290~K & 4.97 & 5.16 & 3.8\% \\
    348~K & 3.96 & 4.24 & 7.1\% \\
    406~K & 2.39 & 2.61 & 9.2\% \\
    464~K & 2.18 & 2.31 & 6.0\% \\
    522~K & 2.07 & 2.19 & 5.8\% \\
    580~K & 2.03 & 2.10 & 3.4\% \\
    638~K & 1.99 & 2.06 & 3.5\% \\
    696~K & 1.96 & 2.01 & 2.6\% \\
    754~K & 1.93 & 1.97 & 2.1\% \\
    812~K & 1.91 & 1.95 & 2.1\% \\
    870~K & 1.90 & 1.93 & 1.6\% \\
    \hline
    \hline
  \end{tabular}
  \label{tab:sigma}
\end{table}
\begin{figure}[h]
  \includegraphics[width=\columnwidth,bb=50 50 410 302]{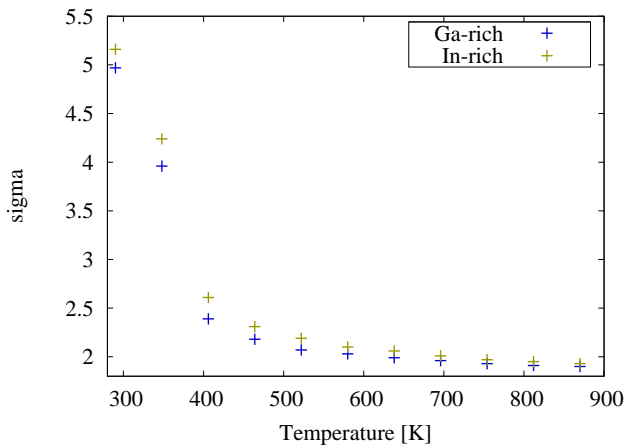}
  \caption{Standard deviation $\sigma$ for In-rich and Ga-rich CIGS as a function of temperature.}
  \label{fig:sigmas}
\end{figure}

\begin{figure}
\begin{tabular}{l}
a)\\
  \begin{minipage}{\columnwidth}
  \includegraphics[width=\columnwidth,bb=50 50 410 302]{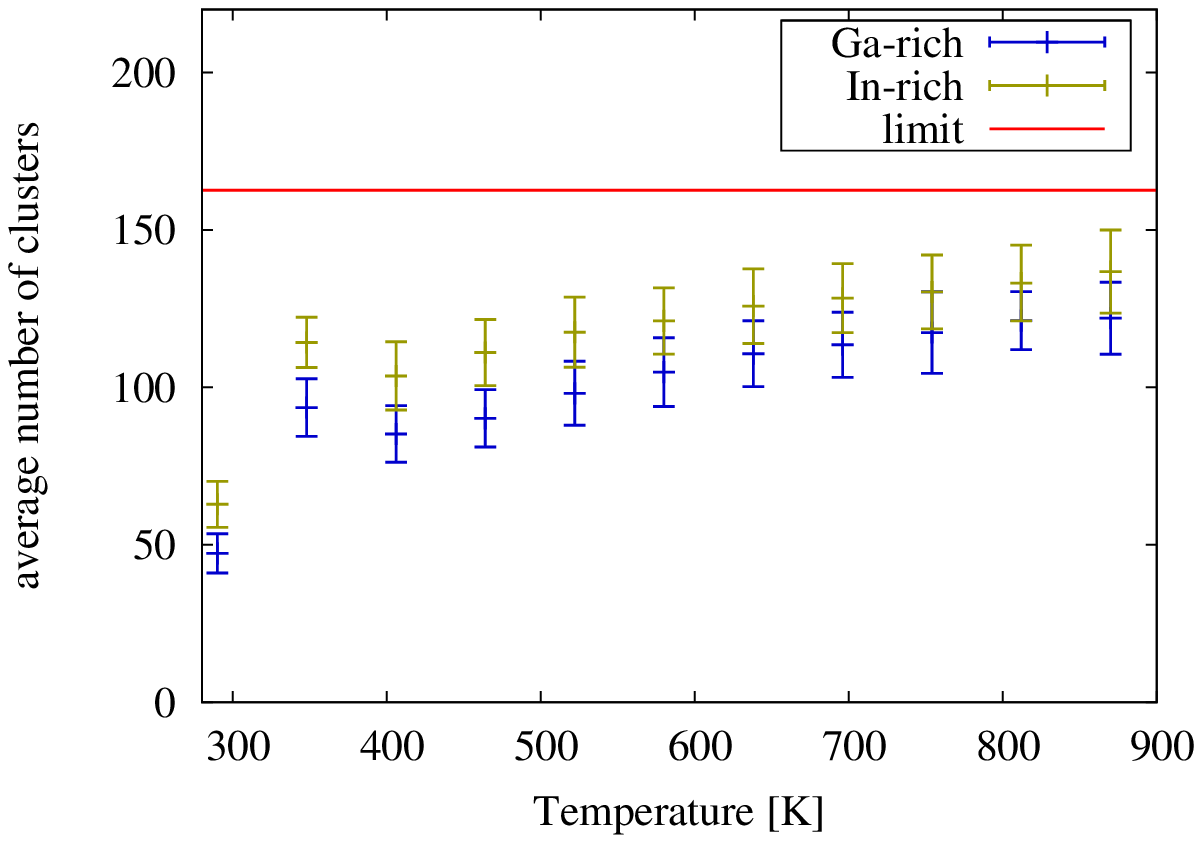}
  \end{minipage}\\
b)\\
  \begin{minipage}[t]{\columnwidth}
    \includegraphics[width=\columnwidth,bb=50 50 410 302]{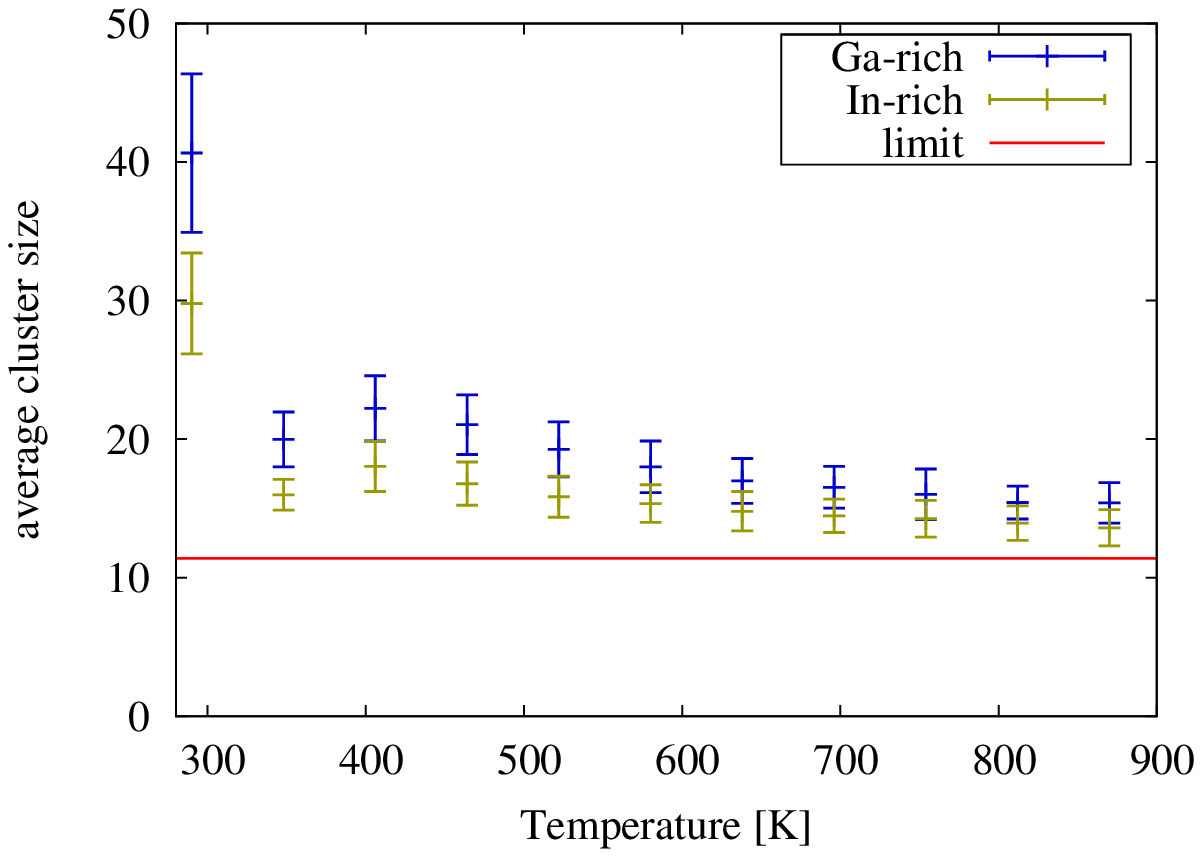}
  \end{minipage}
\end{tabular}
  \caption{Average number of clusters (a) and size of clusters (b) for Ga-rich and In-rich CIGS. Data points are the average value of all data sets of an MC run and error bars are the standard deviation. The blue lines give the limit for simulations with infinite temperature. The considered clusters are connected In atoms in Ga-rich CIGS and connected Ga atoms in In-rich CIGS.}
  \label{fig:domains}
\end{figure}

We have studied a system of CuIn$_{0.25}$Ga$_{0.75}$Se$_2$, which is in the following denoted as {\it Ga-rich} CIGS, and a system of CuIn$_{0.75}$Ga$_{0.25}$Se$_2$, wich is denoted as {\it In-rich} CIGS. Figure \ref{fig:hist} shows histograms of the number of In (Ga) atoms in the segments for Ga-rich and In-rich CIGS. The temperature was kept constant at 25~meV (290~K) and 35~meV (406~K) respectively.

At 25~meV (290~K) the histograms have two maxima: a big peak to the left and a small peak to the right. The majority of the segments contain very few or no In (Ga) atoms, but the small peak indicates that a certain fraction of segments contain a lot of or exclusively In (Ga) atoms. This means there are two phases: an In phase and a Ga phase.
Both peaks are higher for Ga-rich CIGS. Close to the mean value of 4 the values for In-rich CIGS are higher. The standard deviation $\sigma$ is 3.8\% higher for Ga-rich CIGS, indicating a higher inhomogeneity (Tab. \ref{tab:sigma}).

At a temperature of 35~meV (406~K) the system has undergone a phase transition to a mixed, disordered phase. The histograms have changed drastically and show one broad peak with a long tail to the right. This is accompanied by a big change of $\sigma$ to smaller values for both systems (Tab. \ref{tab:sigma}).
The difference of the homogeneity is very pronounced; $\sigma$ is 9.2\% higher for Ga-rich CIGS, the largest difference for all considered temperatures (Fig. \ref{fig:sigmas}).

At higher temperatures the shape of both histograms becomes narrower and the difference between In-rich and Ga-rich CIGS becomes smaller (Fig. \ref{fig:sigmas}).

Table \ref{tab:sigma} contains the $\sigma$ values for more temperatures, for which no histograms are shown. It can be seen that $\sigma$ is smaller for In-rich CIGS at all temperatures and $\sigma$ decreases with temperature for both systems.
The relative difference between In-rich and Ga-rich CIGS is largest at 35~meV (406~K). The focus of this paper is not on the phase transition and we did not determine the exact transition temperature. Figures \ref{fig:snapshots}c) and \ref{fig:snapshots}d) show snapshots of a demixed state at 30~meV (348~K) and a mixed state at 35~meV (406~K), respectively. Both simulations started with an ordered system of periodic unit cells (Fig. \ref{fig:snapshots}b))

For further analysis we define clusters of In (Ga) as a number of joint In (Ga) atoms. A low number of clusters with a high average cluster size is a sign for high inhomogeneity.
We look at clusters of the minority atom species; Ga in In-rich CIGS and In in Ga-rich CIGS.
Figure \ref{fig:domains} shows the average number of clusters and cluster size for In-rich and Ga-rich CIGS. Data were taken at several temperatures between 25~meV (290~K) and 200~meV (2321~K). At all temperatures the number of clusters is higher and the size of clusters is lower for In-rich CIGS, confirming the fact that Ga-rich CIGS is more inhomogeneous. The data show a continuous increase of the average number of clusters with temperature for both systems, apart from a small peak at 30~meV (near the phase transition). The increase is rapid below 30~meV and slower above 35~meV. The average size of clusters shows the opposite trend: rapid decrease below 30~meV, a dip at 30~meV and a slower decrease above 35~meV. The horizontal lines in both graphs mark the limits that were obtained in simulations with infinite temperature.

Calculations with bigger simulation boxes (24x24x12) show that finite size effects do not play a role for these results. The size of the clusters is independent of the volume, as is the ratio number of clusters : volume.

\bigskip

We investigated the homogeneity of CuIn$_{x}$Ga$_{1-x}$Se$_2$ at different temperatures. By studying the spatial distribution and joint clusters of In and Ga we showed that In-rich CIGS exhibits a higher homogeneity than Ga-rich CIGS at all considered temperatures between room temperature and the approximate production temperature of solar cells. This is in agreement with the experiments of G\"utay and Bauer \cite{bauer_pl1}. The effect of cluster size dependence on Ga content 
provides a possible explanation for the relatively low efficiency of CIGS 
with high Ga content (low as compared to what could be excepted from their 
band-gap in the homogeneous case). 

Our results show that inhomogeneities get strongly pronounced as the material is cooled down to room temperature, undergoing the demixing transition. This means that a much more homogeneous sample should be expected if one would ``freeze in'' the high-temperature state by a faster cooling of the material. This prediction should be verifiable experimentally and it might have a valuable impact on the solar cell efficiency.

Above the demixing temperature the size of the clusters is in a regime where simple coexistence of two phases is not an adequate description. A new approach will be necessary to accurately calculate electronic properties of the material.

\bigskip
\begin{acknowledgments}
The authors would like to thank Axel van de Walle for help with the \footnotesize ATAT \normalsize program package.

This work was funded by the German \mbox{Bundesministerium} f{\"{u}}r Umwelt, Naturschutz und Reaktorsicherheit (project 0327665A).
\end{acknowledgments}



\begin{thebibliography}{19}
\expandafter\ifx\csname natexlab\endcsname\relax\def\natexlab#1{#1}\fi
\expandafter\ifx\csname bibnamefont\endcsname\relax
  \def\bibnamefont#1{#1}\fi
\expandafter\ifx\csname bibfnamefont\endcsname\relax
  \def\bibfnamefont#1{#1}\fi
\expandafter\ifx\csname citenamefont\endcsname\relax
  \def\citenamefont#1{#1}\fi
\expandafter\ifx\csname url\endcsname\relax
  \def\url#1{\texttt{#1}}\fi
\expandafter\ifx\csname urlprefix\endcsname\relax\def\urlprefix{URL }\fi
\providecommand{\bibinfo}[2]{#2}
\providecommand{\eprint}[2][]{\url{#2}}

\bibitem[{\citenamefont{Huang}(2008)}]{huang_effects_of_Ga}
\bibinfo{author}{\bibfnamefont{C.-H.} \bibnamefont{Huang}},
  \bibinfo{journal}{J. of Phys. and Chem. of Solids}
  \textbf{\bibinfo{volume}{69}}, \bibinfo{pages}{330} (\bibinfo{year}{2008}).

\bibitem[{\citenamefont{Su-Huai~Wei and Zunger}(1998)}]{zunger_effects_of_Ga}
\bibinfo{author}{\bibfnamefont{S.~Z.} \bibnamefont{Su-Huai~Wei}}
  \bibnamefont{and} \bibinfo{author}{\bibfnamefont{A.}~\bibnamefont{Zunger}},
  \bibinfo{journal}{Appl. Phys. Lett.} \textbf{\bibinfo{volume}{72}},
  \bibinfo{pages}{3199} (\bibinfo{year}{1998}).

\bibitem[{\citenamefont{Green et~al.}(2005)\citenamefont{Green, Emery, King,
  Igari, and Warta}}]{record_eff}
\bibinfo{author}{\bibfnamefont{M.}~\bibnamefont{Green}},
  \bibinfo{author}{\bibfnamefont{K.}~\bibnamefont{Emery}},
  \bibinfo{author}{\bibfnamefont{D.}~\bibnamefont{King}},
  \bibinfo{author}{\bibfnamefont{S.}~\bibnamefont{Igari}}, \bibnamefont{and}
  \bibinfo{author}{\bibfnamefont{W.}~\bibnamefont{Warta}},
  \bibinfo{journal}{Prog. Photovolt: Res. Appl.} \textbf{\bibinfo{volume}{13}},
  \bibinfo{pages}{49} (\bibinfo{year}{2005}).

\bibitem[{\citenamefont{Werner et~al.}(2005)\citenamefont{Werner, Mattheis, and
  Rau}}]{eff_limit}
\bibinfo{author}{\bibfnamefont{J.}~\bibnamefont{Werner}},
  \bibinfo{author}{\bibfnamefont{J.}~\bibnamefont{Mattheis}}, \bibnamefont{and}
  \bibinfo{author}{\bibfnamefont{U.}~\bibnamefont{Rau}}, \bibinfo{journal}{Thin
  Solid Films} \textbf{\bibinfo{volume}{480}}, \bibinfo{pages}{399}
  (\bibinfo{year}{2005}).

\bibitem[{\citenamefont{G{\"{u}}tay and Bauer}(2007)}]{bauer_pl1}
\bibinfo{author}{\bibfnamefont{L.}~\bibnamefont{G{\"{u}}tay}} \bibnamefont{and}
  \bibinfo{author}{\bibfnamefont{G.}~\bibnamefont{Bauer}},
  \bibinfo{journal}{Thin Solid Films} \textbf{\bibinfo{volume}{515}},
  \bibinfo{pages}{6212} (\bibinfo{year}{2007}).

\bibitem[{\citenamefont{G{\"{u}}tay and Bauer}(2009)}]{bauer_pl2}
\bibinfo{author}{\bibfnamefont{L.}~\bibnamefont{G{\"{u}}tay}} \bibnamefont{and}
  \bibinfo{author}{\bibfnamefont{G.}~\bibnamefont{Bauer}},
  \bibinfo{journal}{Thin Solid Films} \textbf{\bibinfo{volume}{517}},
  \bibinfo{pages}{2222} (\bibinfo{year}{2009}).

\bibitem[{\citenamefont{G{\"{u}}tay et~al.}(2009)\citenamefont{G{\"{u}}tay,
  Pomraenke, Lienau, and Bauer}}]{bauer_pl3}
\bibinfo{author}{\bibfnamefont{L.}~\bibnamefont{G{\"{u}}tay}},
  \bibinfo{author}{\bibfnamefont{R.}~\bibnamefont{Pomraenke}},
  \bibinfo{author}{\bibfnamefont{C.}~\bibnamefont{Lienau}}, \bibnamefont{and}
  \bibinfo{author}{\bibfnamefont{G.}~\bibnamefont{Bauer}},
  \bibinfo{journal}{Phys. Status Solidi A} \textbf{\bibinfo{volume}{206}},
  \bibinfo{pages}{1005} (\bibinfo{year}{2009}).

\bibitem[{\citenamefont{Yan et~al.}(2005)\citenamefont{Yan, Noufi, Jones,
  Ramanathan, Al-Jassim, and Stanbery}}]{inhom_nano}
\bibinfo{author}{\bibfnamefont{Y.}~\bibnamefont{Yan}},
  \bibinfo{author}{\bibfnamefont{R.}~\bibnamefont{Noufi}},
  \bibinfo{author}{\bibfnamefont{K.}~\bibnamefont{Jones}},
  \bibinfo{author}{\bibfnamefont{K.}~\bibnamefont{Ramanathan}},
  \bibinfo{author}{\bibfnamefont{M.}~\bibnamefont{Al-Jassim}},
  \bibnamefont{and} \bibinfo{author}{\bibfnamefont{B.}~\bibnamefont{Stanbery}},
  \bibinfo{journal}{Appl. Phys. Lett.} \textbf{\bibinfo{volume}{87}},
  \bibinfo{pages}{121904} (\bibinfo{year}{2005}).

\bibitem[{\citenamefont{Rega et~al.}(2005)\citenamefont{Rega, Siebentritt,
  Albert, Nishiwaki, Zajogin, Lux-Steiner, Kniese, and Romero}}]{pl_vary_Ga}
\bibinfo{author}{\bibfnamefont{N.}~\bibnamefont{Rega}},
  \bibinfo{author}{\bibfnamefont{S.}~\bibnamefont{Siebentritt}},
  \bibinfo{author}{\bibfnamefont{J.}~\bibnamefont{Albert}},
  \bibinfo{author}{\bibfnamefont{S.}~\bibnamefont{Nishiwaki}},
  \bibinfo{author}{\bibfnamefont{A.}~\bibnamefont{Zajogin}},
  \bibinfo{author}{\bibfnamefont{M.}~\bibnamefont{Lux-Steiner}},
  \bibinfo{author}{\bibfnamefont{R.}~\bibnamefont{Kniese}}, \bibnamefont{and}
  \bibinfo{author}{\bibfnamefont{M.}~\bibnamefont{Romero}},
  \bibinfo{journal}{Thin Solid Films} \textbf{\bibinfo{volume}{480}},
  \bibinfo{pages}{286} (\bibinfo{year}{2005}).

\bibitem[{\citenamefont{Jaffe and
  Zunger}(1983)}]{zunger_electronic_structure_of}
\bibinfo{author}{\bibfnamefont{J.~E.}~\bibnamefont{Jaffe}} \bibnamefont{and}
  \bibinfo{author}{\bibfnamefont{A.}~\bibnamefont{Zunger}},
  \bibinfo{journal}{Phys. Rev. B} \textbf{\bibinfo{volume}{28}},
  \bibinfo{pages}{5822} (\bibinfo{year}{1983}).

\bibitem[{\citenamefont{Sanchez et~al.}(1984)\citenamefont{Sanchez, Ducastelle,
  and Gratias}}]{ce_basics}
\bibinfo{author}{\bibfnamefont{J.}~\bibnamefont{Sanchez}},
  \bibinfo{author}{\bibfnamefont{F.}~\bibnamefont{Ducastelle}},
  \bibnamefont{and} \bibinfo{author}{\bibfnamefont{D.}~\bibnamefont{Gratias}},
  \bibinfo{journal}{Physica A} \textbf{\bibinfo{volume}{128}},
  \bibinfo{pages}{334} (\bibinfo{year}{1984}).

\bibitem[{\citenamefont{M{\"{u}}ller}(2003)}]{ce_basics2}
\bibinfo{author}{\bibfnamefont{S.}~\bibnamefont{M{\"{u}}ller}},
  \bibinfo{journal}{J. Phys. Condens. Matter} \textbf{\bibinfo{volume}{15}},
  \bibinfo{pages}{R1429} (\bibinfo{year}{2003}).

\bibitem[{abi()}]{abinit_note}
\emph{\bibinfo{title}{The abinit code is a common project of the université
  catholique de louvain, corning incorporated, and other contributors (url
  http://www.abinit.org).}}

\bibitem[{\citenamefont{Gonze et~al.}(2002)\citenamefont{Gonze, Beuken,
  Caracas, Detraux, Fuchs, Rignanese, Sindic, Verstraete, Zerah, Jollet
  et~al.}}]{abinit}
\bibinfo{author}{\bibfnamefont{X.}~\bibnamefont{Gonze}},
  \bibinfo{author}{\bibfnamefont{J.-M.} \bibnamefont{Beuken}},
  \bibinfo{author}{\bibfnamefont{R.}~\bibnamefont{Caracas}},
  \bibinfo{author}{\bibfnamefont{F.}~\bibnamefont{Detraux}},
  \bibinfo{author}{\bibfnamefont{M.}~\bibnamefont{Fuchs}},
  \bibinfo{author}{\bibfnamefont{G.-M.} \bibnamefont{Rignanese}},
  \bibinfo{author}{\bibfnamefont{L.}~\bibnamefont{Sindic}},
  \bibinfo{author}{\bibfnamefont{M.}~\bibnamefont{Verstraete}},
  \bibinfo{author}{\bibfnamefont{G.}~\bibnamefont{Zerah}},
  \bibinfo{author}{\bibfnamefont{F.}~\bibnamefont{Jollet}},
  \bibnamefont{et~al.}, \bibinfo{journal}{Computational Materials Science}
  \textbf{\bibinfo{volume}{25}}, \bibinfo{pages}{478} (\bibinfo{year}{2002}).

\bibitem[{\citenamefont{van~de Walle and Ceder}(2002)}]{atat}
\bibinfo{author}{\bibfnamefont{A.}~\bibnamefont{van~de Walle}}
  \bibnamefont{and} \bibinfo{author}{\bibfnamefont{G.}~\bibnamefont{Ceder}},
  \bibinfo{journal}{J. of Phase Equilibria} \textbf{\bibinfo{volume}{23}},
  \bibinfo{pages}{348} (\bibinfo{year}{2002}).

\bibitem[{\citenamefont{van~de Walle et~al.}(2002)\citenamefont{van~de Walle,
  Asta, and Ceder}}]{atat_userguide}
\bibinfo{author}{\bibfnamefont{A.}~\bibnamefont{van~de Walle}},
  \bibinfo{author}{\bibfnamefont{M.}~\bibnamefont{Asta}}, \bibnamefont{and}
  \bibinfo{author}{\bibfnamefont{G.}~\bibnamefont{Ceder}},
  \bibinfo{journal}{CALPHAD Journal} \textbf{\bibinfo{volume}{26}},
  \bibinfo{pages}{539} (\bibinfo{year}{2002}).

\bibitem[{\citenamefont{Perdew et~al.}(1996)\citenamefont{Perdew, Burke, and
  Ernzerhof}}]{pbe-gga}
\bibinfo{author}{\bibfnamefont{J.~P.} \bibnamefont{Perdew}},
  \bibinfo{author}{\bibfnamefont{K.}~\bibnamefont{Burke}}, \bibnamefont{and}
  \bibinfo{author}{\bibfnamefont{M.}~\bibnamefont{Ernzerhof}},
  \bibinfo{journal}{Phys. Rev. Lett.} \textbf{\bibinfo{volume}{77}},
  \bibinfo{pages}{3865} (\bibinfo{year}{1996}).

\bibitem[{\citenamefont{H{\"{u}}lsen et~al.}(2009)\citenamefont{H{\"{u}}lsen,
  Scheffler, and Kratzer}}]{kratzer_ce}
\bibinfo{author}{\bibfnamefont{B.}~\bibnamefont{H{\"{u}}lsen}},
  \bibinfo{author}{\bibfnamefont{M.}~\bibnamefont{Scheffler}},
  \bibnamefont{and} \bibinfo{author}{\bibfnamefont{P.}~\bibnamefont{Kratzer}},
  \bibinfo{journal}{Phys. Rev. B} \textbf{\bibinfo{volume}{79}},
  \bibinfo{pages}{094407} (\bibinfo{year}{2009}).

\bibitem[{\citenamefont{Laks et~al.}(1992)\citenamefont{Laks, Ferreira, Froyen,
  and Zunger}}]{constituent_strain}
\bibinfo{author}{\bibfnamefont{D.~B.} \bibnamefont{Laks}},
  \bibinfo{author}{\bibfnamefont{L.}~\bibnamefont{Ferreira}},
  \bibinfo{author}{\bibfnamefont{S.}~\bibnamefont{Froyen}}, \bibnamefont{and}
  \bibinfo{author}{\bibfnamefont{A.}~\bibnamefont{Zunger}},
  \bibinfo{journal}{Phys. Rev. B} \textbf{\bibinfo{volume}{46}},
  \bibinfo{pages}{587} (\bibinfo{year}{1992}).

\end{thebibliography}
\end{document}